\documentclass[a4paper,fleqn,usenatbib]{mnras}

\usepackage{newtxtext,newtxmath}

\usepackage[T1]{fontenc}
\usepackage{ae,aecompl}

\usepackage{graphicx}	
\usepackage{amsmath}	
\usepackage{amssymb}	
\usepackage{subfig}
\usepackage{threeparttable}
\usepackage{pdflscape}

\title[Origin of Cosmic Ray Electrons and Positrons]{Origin of Cosmic Ray Electrons and Positrons}

\author[Zhao-Dong Shi and Siming Liu]{
Zhao-Dong Shi$^{1,2}$
\ and Siming Liu$^{1,2}$\thanks{E-mail: liusm@pmo.ac.cn}
\\
$^{1}$Key Laboratory of Dark Matter and Space Astronomy, Purple Mountain Observatory, Chinese Academy of Sciences, Nanjing 210008, Jiangsu, China\\
$^{2}$School of Astronomy and Space Science, University of Science and Technology of China, Hefei 230026, Anhui, China
}

\date{Accepted 2019 March 4. Received 2019 February 25; in original form 2018 December 4}

\pubyear{?}

\begin{document}
\label{firstpage}
\pagerange{\pageref{firstpage}--\pageref{lastpage}}
\maketitle

\begin{abstract}
With experimental results of AMS on the spectra of cosmic ray (CR) $e^{-}$, $e^{+}$, $e^{-}+e^{+}$ and positron fraction, as well as new measurements of CR $e^{-}+e^{+}$ flux by HESS, one can better understand the CR lepton ($e^{-}$ and $e^{+}$) spectra and the puzzling electron-positron excess above $\sim$10 GeV. In this article, spectra of CR $e^{-}$ and $e^{+}$ are fitted with a physically motivated simple model, and their injection spectra are obtained with a one-dimensional propagation model including the diffusion and energy loss processes. Our results show that the electron-positron excess can be attributed to uniformly distributed sources that continuously inject into the galactic disk electron-positron with a power-law spectrum cutting off near 1 TeV and a triple power-law model is needed to fit the primary CR electron spectrum. The lower energy spectral break can be attributed to propagation effects giving rise to a broken power-law injection spectrum of primary CR electrons with a spectral hardening above $\sim$40 GeV.
\end{abstract}

\begin{keywords}
cosmic rays: positrons -- 
diffusion -- injection
\end{keywords}



\section{Introduction} \label{sec:intro}

Measurements of cosmic ray (CR) electrons (${e^{-}}$) and positrons (${e^{+}}$) have been advanced significantly during the past few years. In particular, results of the Advanced Thin Ionization Calorimeter (ATIC) balloon experiment \citep{2008Natur.456..362C} reveal a spectral bump of CR electrons plus positrons (${e^{-} + e^{+}}$) at energies of $\sim$300--800 GeV, and results of the Payload for Antimatter Matter Exploration and Light-nuclei Astrophysics (PAMELA) satellite \citep{2009Natur.458..607A} reveal an anomalous increase of CR positron fraction above 10 GeV \citep{2011PhRvL.106t1101A, 2013PhRvL.111h1102A}. These anomalies have been confirmed recently by measurements made with the Alpha Magnetic Spectrometer (AMS) \citep{2014PhRvL.113l1102A, 2014PhRvL.113v1102A, 2014PhRvL.113l1101A}, the Fermi Large Area Telescope (Fermi-LAT) \citep{2017PhRvD..95h2007A}, the Calorimetric Electron Telescope (CALET) \citep{2017PhRvL.119r1101A} and the Dark Matter Particle Explorer (DAMPE) \citep{2017Natur.552...63D} with higher confidence. Fermi-LAT, CALET, and DAMPE have expanded the spectrum of CR ${e^{-} + e^{+}}$ up to the energy of a few TeV.  And at higher energies, measurement \citep{HESS2017} of CR ${e^{-} + e^{+}}$ spectrum from 0.25 TeV to $\sim$20 TeV by High Energy Stereoscopic System (HESS) has just been released. Results of DAMPE and HESS reveal a spectral break near $\sim$1 TeV.

These results are not compatible with the classical cosmic ray model \citep{1998ApJ...493..694M}, where cosmic ray positrons are secondaries produced as primary cosmic ray nuclei propagate in the Galaxy. An injection of primary cosmic ray positron is needed. Some models attribute the excess of CR positrons to annihilation or decay of dark matter particles \citep{2009PhRvL.103c1103B, 2013PhRvD..88b3013C}, which usually requires large reaction cross section for dark matter particles. Others consider astrophysical origin of  the primary CR positrons \citep{2012CEJPh..10....1P}. 

We try to fit the electron and positron spectra with a simple physically motivated model. Given the absence of prominent high-energy spectral features expected by many models with local and/or transient sources of injection \citep{1995A&A...294L..41A, 2004ApJ...601..340K, 2010A&A...524A..51D, 2017ApJ...845..107D, 2017ApJ...836..172F}, we consider the scenario where electron-positron pairs are continuously injected into the whole Galactic disk and a simple 1D propagation model is proposed to obtain the injection spectra. 

The outline of this paper is as follows: a parametrized model for spectra of CR $e^{-}$ and $e^{+}$ is given in Section \ref{sec:spect}. In Section \ref{sec:prop}, we discuss a simple model for CR $e^{-}$ and $e^{+}$ propagation in the Galaxy. The injection spectra of CR $e^{-}$ and $e^{+}$ is  then obtained in Section \ref{sec:inj}. Finally, we draw our conclusions and give some discussions in Section \ref{sec:conc}.

\section{Modeling the spectra of cosmic ray electrons and positrons} \label{sec:spect}

AMS has the most accurate measurement of the spectra of CR $e^{-}$, $e^{+}$, ${e^{-} + e^{+}}$ and the positron fraction below $\sim$1 TeV. HESS measures the $e^{-} + e^{{+}}$ spectrum from sub-TeV to tens of TeV. Since there are systematic differences between AMS data and those from Fermi-LAT and DAMPE, we only fit the latest data from AMS and HESS. We focus on the high-energy parts of spectra ($\gtrsim$ 10 GeV), and fit the low-energy parts of spectra via solar modulation.

\citet{2014PhRvL.113l1101A} fit the AMS positron fraction with a minimal model where the fluxes of CR $e^{-}$ and $e^{+}$ are parametrized as the sum of a power law component and a common power law with an exponential cutoff component. But their model can not fit the spectra of CR $e^{-}$ and $e^{-} + e^{+}$. In fact, two breaks in the spectrum of primary CR $e^{-}$ must be introduced for a satisfactory fit to spectra of CR $e^{-}$ and ${e^{-} + e^{+}}$. Hence, extending the model of \citet{2014PhRvL.113l1101A}, the fluxes of CR $e^{-}$ and $e^{+}$ after propagation in the interstellar medium (ISM) are modeled as follows
\begin{equation}
   \begin{aligned}
      J_{e^{-}}&=J^{\mathrm{s}} + J^{-} + 0.6J^{+} \\
      J_{e^{+}}&=J^{\mathrm{s}} + J^{+},
   \end{aligned} \label{eq:eqfit} \end{equation} \\
where,
\begin{equation}
   \begin{aligned}
      J^{-}&= C_{e^{-}} \begin{cases}
\begin{array}{lr} E^{- \gamma _{e^{-}}^{1}}, & E \leq E_{\mathrm{br}1} \\ E_{\mathrm{br}1}^{\gamma _{e^{-}}^{2} - \gamma _{e^{-}}^{1}}E^{- \gamma _{e^{-}}^{2}}, & E_{\mathrm{br}1} < E \leq E_{\mathrm{br}2} \\ E_{\mathrm{br}1}^{\gamma _{e^{-}}^{2} - \gamma _{e^{-}}^{1}}E_{\mathrm{br}2}^{\gamma _{e^{-}}^{3} - \gamma _{e^{-}}^{2}}E^{-\gamma _{e^{-}}^{3}}, & E > E_{\mathrm{br}2} \end{array} \end{cases} \\
J^{+}&= C_{e^{+}}E^{-\gamma _{e^{+}}} \\
J^{\mathrm{s}}&= C_{\mathrm{s}}E^{-\gamma_{\mathrm{s}}}\exp({-{E / E_{\mathrm{cut}}}}).
   \end{aligned}
   \label{eq:eqcomp}
\end{equation}
The common component $J^{\mathrm{s}}$ of electrons and positrons is modeled as a power law with an exponential cutoff whose normalization is $C_{\mathrm{s}}$, spectral index is $\gamma_{\mathrm{s}}$, and cutoff energy is $E_{\mathrm{cut}}$. The primary electron flux $J^{-}$ is modeled as a tripe power law with two break energies $E_{\mathrm{br1}}$ and $E_{\mathrm{br2}}$ whose normalization is $C_{e^{-}}$ and spectral indices are $\gamma_{e^{-}}^{1}$, $\gamma_{e^{-}}^{2}$ and $\gamma_{e^{-}}^{3}$. Secondary positron flux $J^{+}$ is modeled as a power law whose normalization is $C_{e^{+}}$ and spectral index is $\gamma_{e^{+}}$. Secondary electrons and positrons are produced by the interaction of primary CR nuclei with the ISM. The ratio of secondary electrons to secondary positrons is fixed at 0.6 since secondary positron production is greater than secondary electrons due to conservation of charge for positively charged primary CR nuclei \citep{2006ApJ...647..692K, 2009A&A...501..821D}. In the following, all energies are given in units of GeV.

The fluxes of CR $e^{-}$ and $e^{+}$ are influenced by solar modulation. In the present work, the force-field approximation \citep{1968ApJ...154.1011G} which models the solar modulation via an effective potential $\phi$ is adopted. Since electrons are highly-relativistic in the GeV--TeV range we are considering, the observed flux at the Earth is given by
\begin{equation}
J(E)={\left({E \over E + e\phi}\right)}^{2} J_0(E + e\phi)
\end{equation}
with $e$ the elementary charge and $J_0$ is the flux outside the Heliosphere.

\begin{figure*}
   \centering
   \subfloat[]{\includegraphics[width=0.33\textwidth]{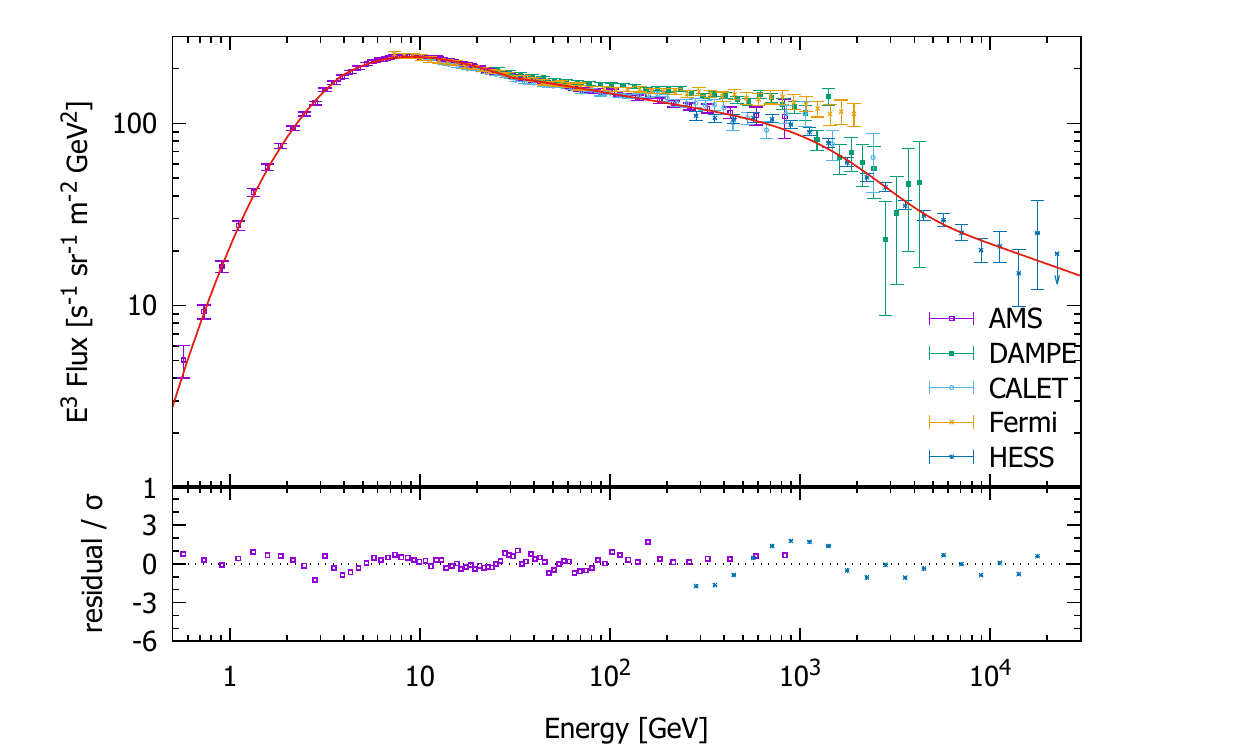}}
   \subfloat[]{\includegraphics[width=0.33\textwidth]{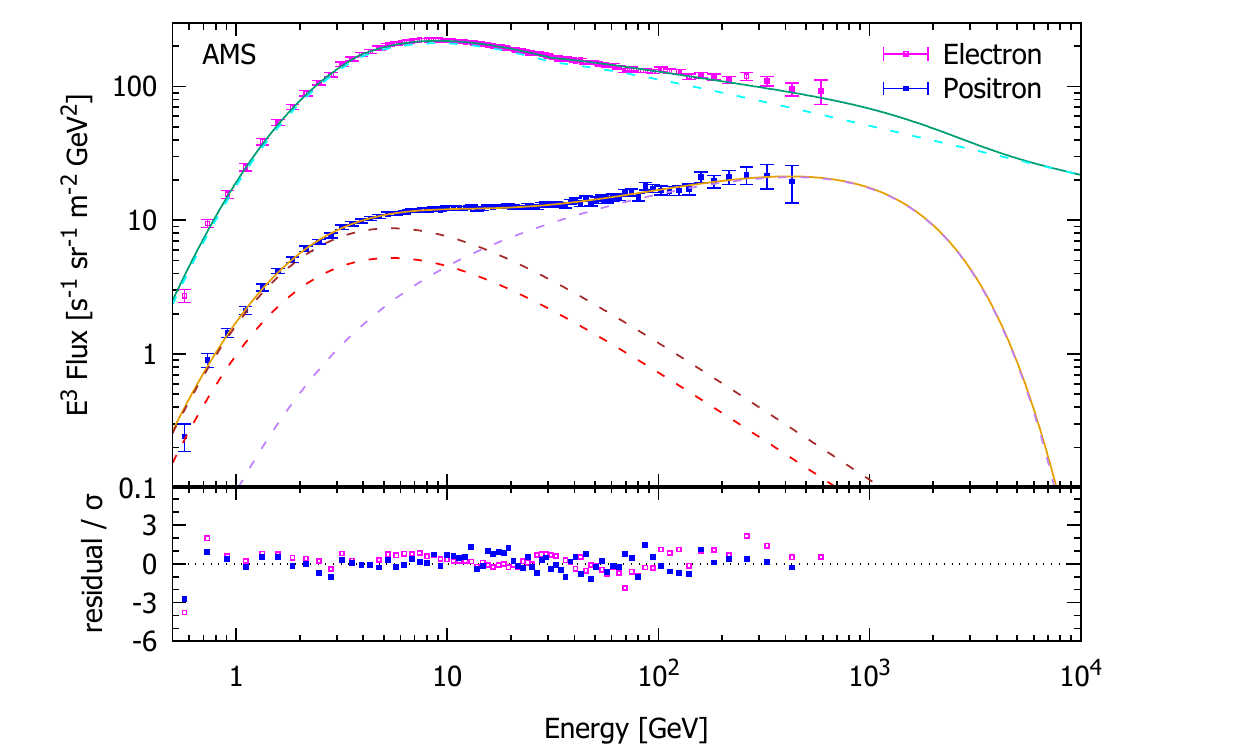}}
   \subfloat[]{\includegraphics[width=0.33\textwidth]{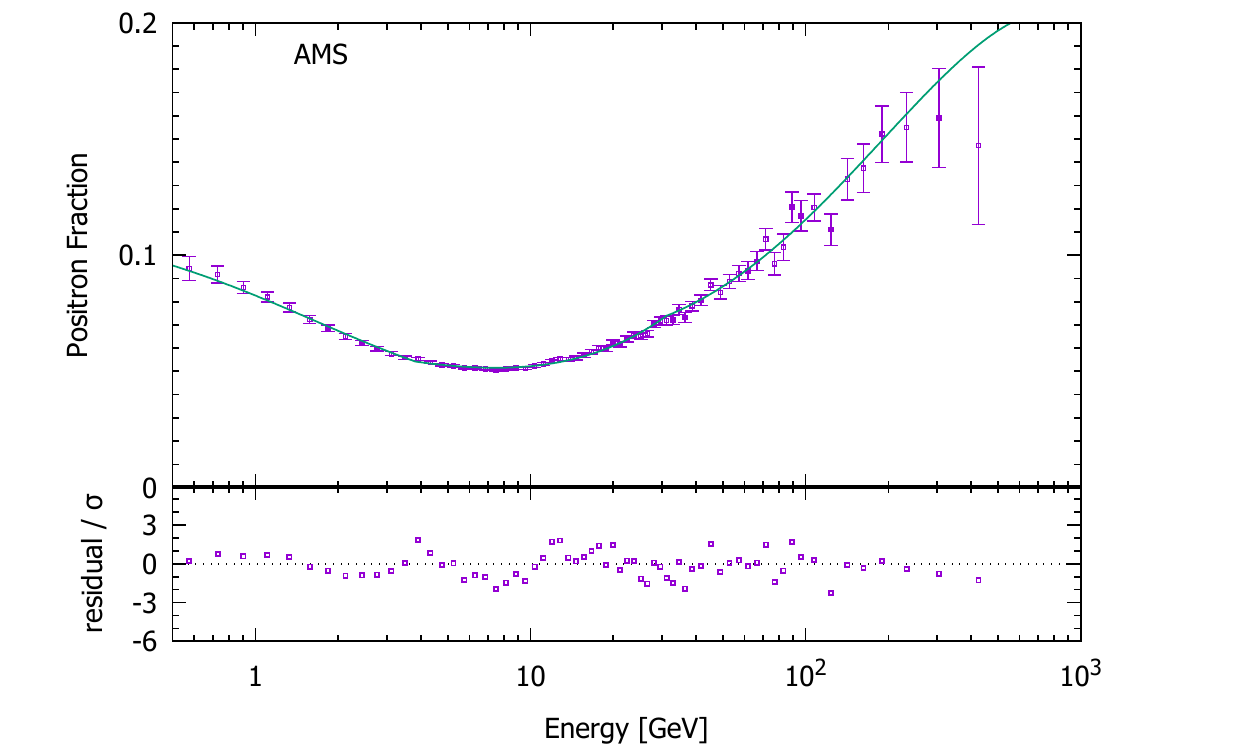}}
  \caption{The fit results according to the model described by Equation (\ref{eq:eqfit}). (a): CR $e^{-} + e^{+}$. (b): CR $e^{-}$ (magenta points) and $e^{+}$ (blue points). The cyan dashed line is the flux of primary $e^{-}$, the brown and red dashed lines are the fluxes of secondary $e^{+}$ and $e^{-}$ respectively, the purple dashed line is the common $e^{-}$ or $e^{+}$ flux. The sea-green and orange lines are the total fluxes of $e^{-}$ and $e^{+}$, respectively. (c): The positron fraction. The ratio of the residual to the standard deviation $\sigma$ of the data are shown in the lower panels (for AMS and HESS data only).}
  \label{fig:fig1}
\end{figure*}

Figure \ref{fig:fig1} shows our best fit (we denote this model by \texttt{M}) to the spectra of $e^{-} + e^{+}$ (a), $e^{-}$ (b), $e^{+}$ (b), and positron fraction (c). The best-fit parameters are listed in Table \ref{tab:pars}. There is no systematic variation in the residuals and the reduced $\chi^{2}_{\mathrm{tot}}$ of our best fit is 0.63 for 304 data points and 12 free parameters (see Table \ref{tab:chisq}). Note that the primary electron spectrum softens near 5 GeV and hardens near 30 GeV and its flux is about 6 times higher than the secondary fluxes at 1 GeV. It is interesting to note that the excesses of CR electrons and positrons, the anomalous increase of positron fraction and the spectral break of HESS data can be explained by the common component $J^{\mathrm{s}}$ with a cutoff energy of $\sim$1 TeV. If such a component has an astrophysical origin, it may experience the same propagation effects as the primary CR electrons $J^{-}$. Without considering details of CR nuclei propagation, the secondary electrons and positrons are subject to the same propagation effects as well. We next construct a simple CR propagation model to obtain the corresponding injection spectra.

\section{The propagation of cosmic ray electrons and positrons in the galaxy} \label{sec:prop}

The propagation of cosmic ray $e^{-}$ and $e^{+}$ in the Galaxy can be described as a diffusion process with significant energy loss \citep[and references therein]{2009PhRvD..80f3005M, 2009A&A...501..821D}. At the presence of large scale magnetic fields, the diffusion coefficient along the magnetic field is more than 10 times larger than the perpendicular diffusion \citep{1999ApJ...520..204G}. One therefore may construct a 1D diffusion model to accout for the CR transport along Galactic magnetic field (GMF) lines \citep{2014Sci...343..988S, 10.1017/S1743921316006530}. The steady-state transport equation for the CR distribution function $N(z, E)$ along GMF is given by
\begin{equation} \label{eq:eqdiffloss}
D(E){\partial ^{2} N \over \partial z^{2}} + {\partial [b(E)N] \over \partial E} + Q(z, E) = 0,
\end{equation}
where the diffusion coefficient $D(E) = D_{0}(E/1 \,\mathrm{GeV})^{\delta}$. For CR $e^{-}$ and $e^{+}$, energy losses are dominated by synchrotron and inverse Compton radiations with the energy loss rate (in the Thompson limit) $\dot{E} \equiv -b(E) = -b_{0}E^{2}$, where \citep{1995PhRvD..52.3265A}
\begin{equation}
b_{0} = 3.21 \times 10^{-6}\,\mathrm{GeV^{-1}\,kyr^{-1}} \cdot {{B^{2}\over8\pi}+ w_{\mathrm{ph}} \over 1\,\mathrm{eV\,cm^{-3}}},
\end{equation}
where $B$ is the GMF and $w_{\mathrm{ph}}$ is the energy density of background photons  including cosmic microwave background (CMB), IR, and starlight. 

The Galaxy is modeled as a disk with half-thickness $h$ and a halo with half-thickness $H>h$ with a vertical magnetic field. Since we ignore the diffusion perpendicular to the magnetic field, the radial structure of the disk is irrelevant. We assume a homogeneous source term $Q(z,E)=q(E)\theta(h-|z|)$, where $\theta(x)$ is Heaviside step function and $q(E)$ is the injection spectrum. We assume no CR at the boundary of the halo so that $N(z=\pm H,E) = 0$.

One can solve Equation (\ref{eq:eqdiffloss}) via Fourier series:
\begin{equation} \label{eq:eqndens}
\begin{aligned}
N&(z,E)={2 \over b_{0}E^{2}} \sum _{m=0}^{\infty} {\sin\left[\left(m+{1\over 2}\right)\pi{h\over H}\right] \over \left(m+{1\over 2}\right)\pi}\cos\left[\left(m+{1\over2}\right)\pi {z\over H}\right] \\ &\times\int _E^{\infty} {\mathrm{d}E{'}q\left(E{'}\right)\exp\left\{{\left[\left(m+{1\over 2}\right) \pi {\ell(E) \over H}\right]}^{2}\left[\left({E\over E{'}}\right)^{1-\delta}-1\right]\right\}}.
\end{aligned}
\end{equation}
\noindent At a given energy $E$, there is characteristic length
\begin{equation}
\ell(E) \equiv \sqrt{D_0E^{\delta -1} \over (1-\delta)b_0},
\end{equation}
where the diffusion timescale $\tau_{\mathrm{diff}} \sim {\ell}^{2} / D(E)$ approximately equals to the energy loss timescale $\tau_{\mathrm{loss}} \sim 1 / b_{0}E$. We can then obtain two characteristic energies $E_{H}$ and $E_{h}$ from the following equations:
\begin{equation}
\ell(E_H)=H,\quad \ell(E_h)=h.
\end{equation}

\begin{figure*}
   \centering
   \subfloat[]{\includegraphics[width=0.25\textwidth]{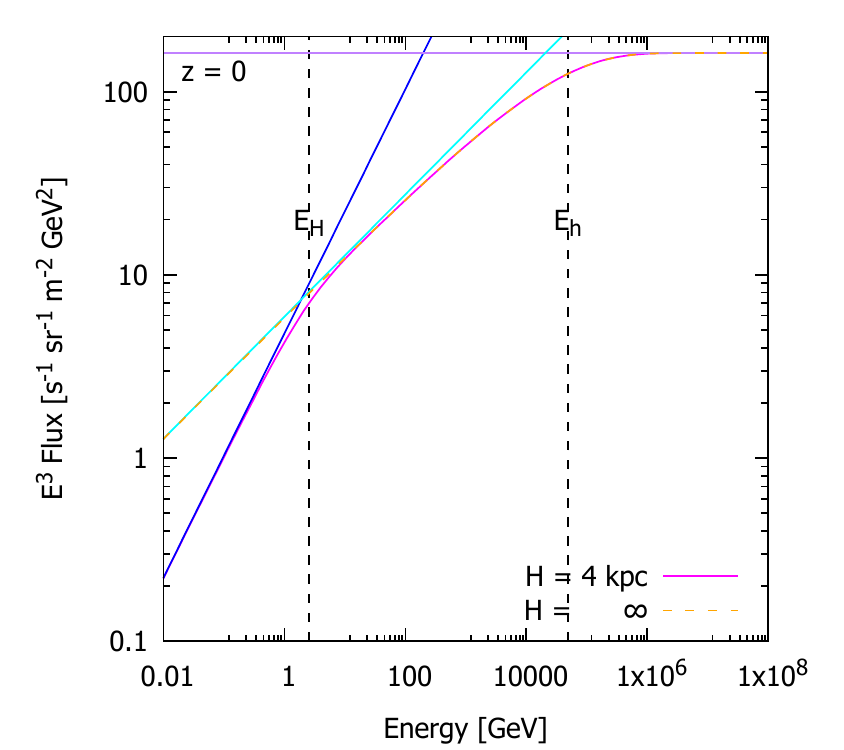}}
   \subfloat[]{\includegraphics[width=0.25\textwidth]{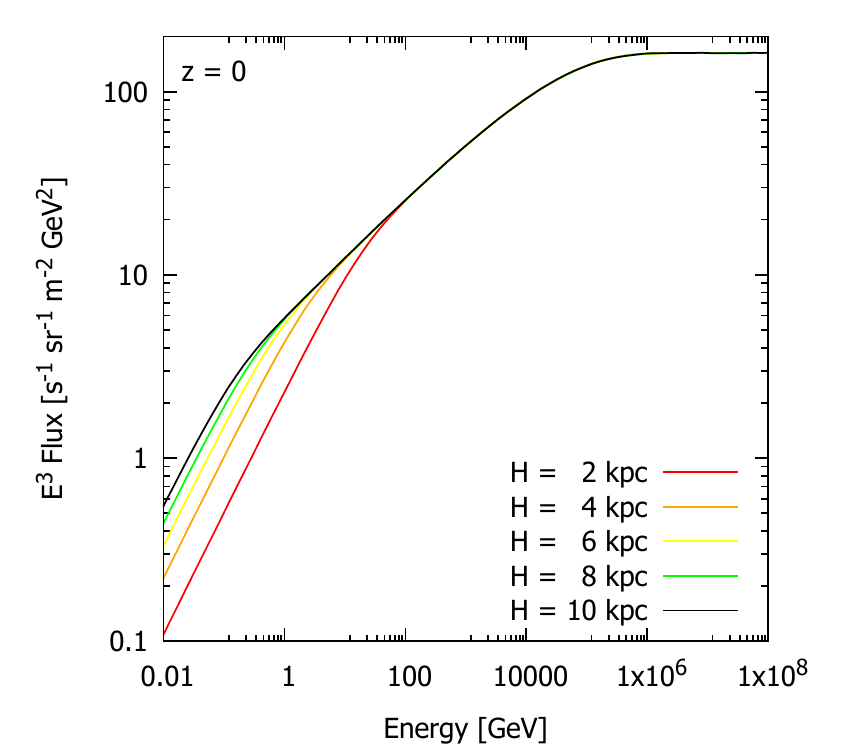}}
   \subfloat[]{\includegraphics[width=0.25\textwidth]{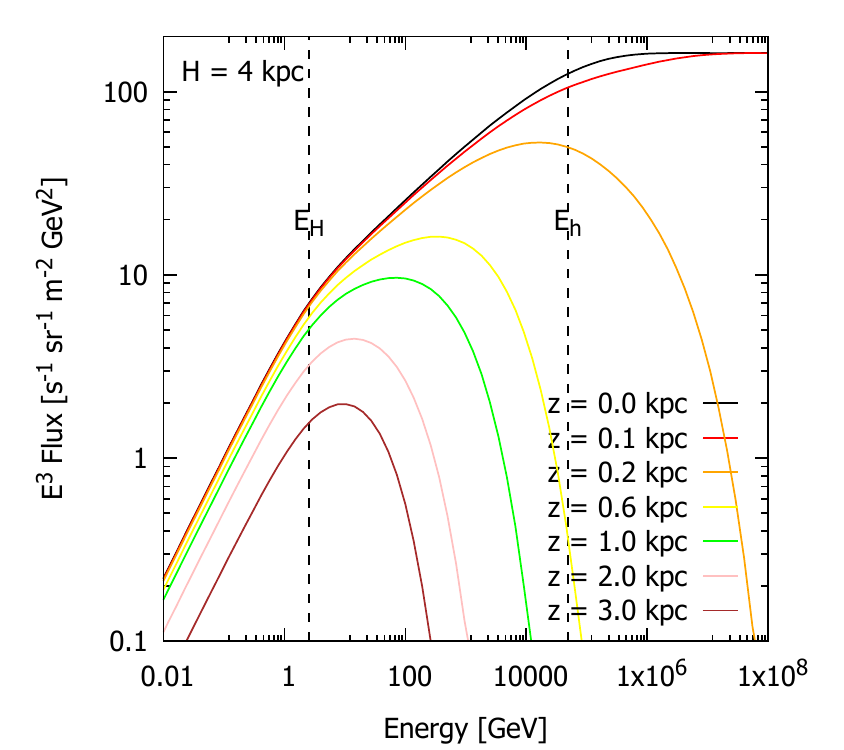}}
   \subfloat[]{\includegraphics[width=0.25\textwidth]{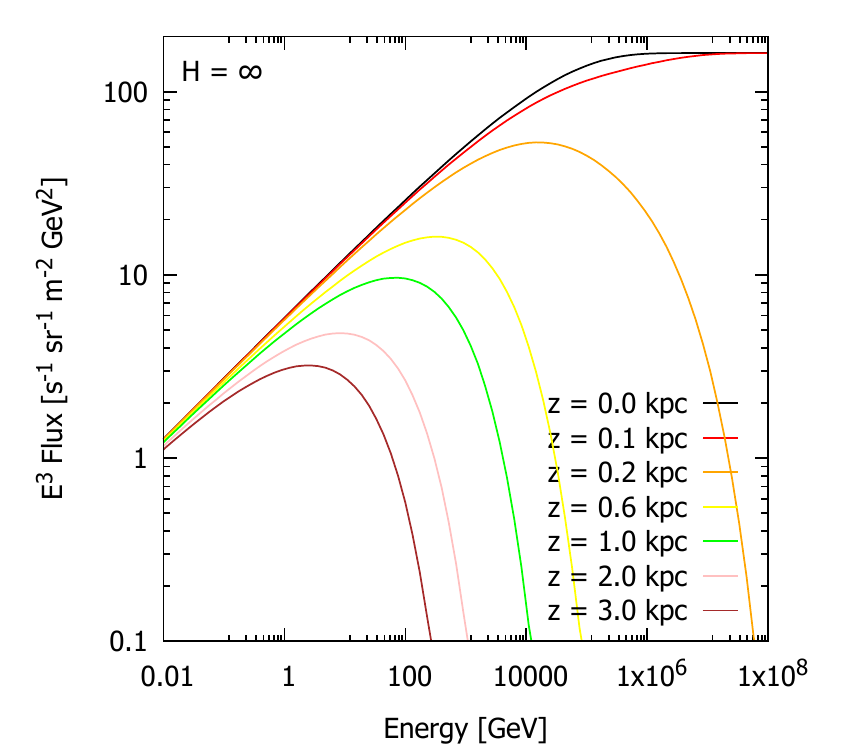}}
   \caption{The spectra after propagation for a power-law injection. The half-thickness of the disk is fixed at $h = 150\,\mathrm{pc}$ and the other common parameters are given in Table \ref{tab:fitprop}. (a): Magenta line is the spectrum for $H=4\,\mathrm{kpc}$, whose asymptotic behaviors (Equation (\ref{eq:eqindex})) are shown as straight blue, cyan, and purple lines with an index $\gamma _a = \gamma + \delta$, $\gamma + {\delta + 1 \over 2}$, $\gamma + 1$, respectively. The orange dashed line is for $H=\infty$. $E_{H}$ and $E_{h}$ are indicated by two vertical lines. (b): Dependence of the spectrum on the galactic plane ($z=0$) on $H$. (c) \& (d): Dependence of the spectrum on $z$ for $H=4\,\mathrm{kpc}$ and $H=\infty$, respectively. \label{fig:fig2}}
\end{figure*}

For a power-law injection $q(E) = KE^{-\gamma}$ with $\gamma > 1$ and normalization $K$, Equation (\ref{eq:eqndens}) gives
\begin{equation} \label{eq:eqpl}
\begin{aligned}
N&(z,E)={2KE^{-(\gamma + 1)} \over (1-\delta)b_0} \sum _{m=0}^{\infty} {\sin\left[\left(m+{1\over 2}\right)\pi{h\over H}\right] \over \left(m+{1\over 2}\right)\pi} \\ &\times \cos\left[\left(m+{1\over2}\right)\pi {z\over H}\right] I\left(\left(m+{1\over 2}\right) \pi {\ell(E) \over H}\right),
\end{aligned}
\end{equation}
where,
\begin{equation}
I(x)=\int_{0}^{1}\mathrm{d}p(1-p)^{\gamma+\delta-2 \over 1-\delta}\exp\left(-px^2\right).
\end{equation}
Integral $I(x)$ has the following asymptotic behavior:
\begin{equation} 
I(x) =
\begin{cases}
\frac{1-\delta}{\gamma-1},\quad &x \to 0 \\
1/x^{2},\quad &x \to \infty.
\end{cases}
\end{equation}
It is easy to see that the above asymptotic formulae are applicable to $E<E_{H}$ (i.e., $\ell(E)/H \gg 1$) and $E>E_{h}$ (i.e., $\ell(E)/H \ll 1$). For $E_{H} < E < E_{h}$, there is no asymptotic formula. Fortunately, one may consider the case with $H=\infty$ so that $E_{H}=0$. Then Equation (\ref{eq:eqdiffloss}) can be solved through the Green's function (the solution is denoted by $N_{\infty}$):
\begin{equation} \label{eq:eqinfty}
N_{\infty}(z,E) = {1 \over b(E)} \int_{E}^{\infty} \mathrm{d}E_{0}q(E_{0}) {1 \over \sqrt\pi} \int_{-h \over 2 \sqrt\lambda}^{h \over 2 \sqrt\lambda} \mathrm{d}\xi e^{-\left(\xi - {z \over 2 \sqrt\lambda}\right)^2},
\end{equation}
where,
\begin{equation}
\lambda=\int_E^{E_{0}} {D(E{'}) \over b(E{'})} \mathrm{d}E{'}=\ell^{2}(E)\left[1 - {\left({E\over E_0}\right)}^{1-\delta}\right].
\end{equation}
And for the same power-law injection,
\begin{equation} \label{eq:eqinftypl}
\left. \begin{array}{ll}
&N_{\infty}(z,E)= {hKE^{-\left(\gamma+{\delta+1\over2}\right)}\over \sqrt{\pi(1-\delta)b_{0}D_{0}}} \int_{0}^{1}\mathrm{d}p(1-p^{2})^{\gamma+\delta-2\over1-\delta} \\ &\quad \times\int_{-1}^{1}\mathrm{d}\xi\exp\left\{-\left[{h\over2p\ell(E)}\left(\xi-{z\over h}\right)\right]^{2}\right\}.
\end{array} \right.
\end{equation}
The CR flux is related to $N$ via $J_{0}(z,\,E)=\frac{c}{4\pi}N(z,\,E)$ where $c$ is the speed of light. Then we have\footnote{Two identities have been invoked for $0<x<2$:
\[ \left. \begin{array}{ll} \sum _{m=0}^{\infty}{\sin\left[\left(m+\frac{1}{2}\right)\pi x\right] \over \left(m + \frac{1}{2}\right)\pi} = \frac{1}{2} \\
\sum _{m=0}^{\infty}{\sin\left[\left(m+\frac{1}{2}\right)\pi x\right] \over \left[\left(m + \frac{1}{2}\right)\pi\right]^{3}} = \frac{x}{2}\left(1-\frac{x}{2}\right). \end{array} \right.
\]}
\begin{equation} 
J_{0}(0,\,E) \simeq
\begin{cases}
{cHhKE^{-(\gamma+\delta)}\over4\pi D_{0}}\left(1-\frac{h}{2H}\right),\quad &E<E_{H} \\
{cKE^{-(\gamma+1)}\over4\pi(\gamma-1)b_{0}},\quad &E>E_{h}
\end{cases}
\end{equation}
and for $E_{H}<E<E_{h}$,
\begin{equation}
J_{0}(0,\,E) \simeq {chKE^{-\left(\gamma+{\delta+1\over2}\right)}\over 4\pi\sqrt{(1-\delta)b_{0}D_{0}}}{\Gamma\left(\frac{\gamma-1}{1-\delta}\right)\over \Gamma \left(\frac{\gamma-1}{1-\delta}+\frac{1}{2}\right)}.
\end{equation}
The corresponding spectral index of $J_{0}$ is given by
\begin{equation} \label{eq:eqindex}
\gamma_{a} = 
\begin{cases}
\gamma + \delta,\quad & E < E_{H} \\
\gamma + \frac{\delta + 1}{2},\quad & E_{H} < E < E_{h}\\
\gamma + 1,\quad & E > E_{h}.
\end{cases}
\end{equation}
For the parameters \{$D_{0}$, $\delta$, $b_{0}$, $\gamma$, $K$\} that we choose and are listed in Table \ref{tab:fitprop}, Figure \ref{fig:fig2} shows some of the results and its dependence on $z$ and $H$.

These indexes can be understood qualitatively \citep{1974Ap&SS..29..305B}. The number density $N$ after propagation can be estimated as $N \sim q(E)Th/L$, where $T$ and $L$ are the relevant time and length scales, respectively. For $E<E_{H}$, the diffusion term dominates so that $L\sim H$ and $T\sim H^{2}/D(E)$, hence, $N \sim Hhq(E)/D(E)\propto E^{-(\gamma+\delta)}$. For $E_{H}<E<E_{h}$, $L \sim \ell(E)$ and $T\sim \ell^{2}(E)/D(E)$, hence, $N \sim hq(E)\ell(E)/D(E)\propto E^{-(\gamma+(\delta+1)/2)}$. For $E>E_{h}$, the energy loss dominates so that $L\sim h$ and $T\sim 1/b_{0}E$, hence, $N \sim q(E)/b_{0}E\propto E^{-(\gamma+1)}$.

\begin{landscape}
\begin{table}
\centering
\caption{The best-fit parameters of our phenomenological model \texttt{M} and propagation model \texttt{P1} and \texttt{P2}. While model \texttt{M} is described in Section \ref{sec:spect}, both model \texttt{P1} with one effective potential and model \texttt{P2} with two effective potentials are described in Section \ref{sec:inj}. \label{tab:pars}}
\begin{threeparttable}
\renewcommand\arraystretch{1.8}
\begin{tabular}{*{18}{c}}
\hline
Model & $C_{e^{-}}$\tnote{a} & $\gamma _{e^{-}}^{1}$ & $E_{\mathrm{br}1}$ & $\gamma _{e^{-}}^{2}$ & $E_{\mathrm{br}2}$ & $\gamma _{e^{-}}^{3}$ & $C_{e^{+}}$\tnote{a} & $\gamma _{e^{+}}$ & $C_{\mathrm{s}}$\tnote{a} & $\gamma_{\mathrm{s}}$ & $E_{\mathrm{cut}}$ & & & & & $\phi$\tnote{f} \\
\cline{2-18}
\texttt{M} & $1.06\times10^{3}$ & 3.35 & 4.96 & 3.64 & 32.4 & 3.37 & 163 & 4.05 & 3.14 & 2.62 & $1.10\times10^{3}$ & & & & & 1.14 \\
\hline
Model & $C^{\mathrm{inj}}_{e^{-}}$\tnote{b} & $\gamma_{e^{-}}^{1,\mathrm{inj}}$ & & & $E_{\mathrm{br}}^{\mathrm{inj}}$ & $\gamma_{e^{-}}^{2,\mathrm{inj}}$ & $C^{\mathrm{inj}}_{e^{+}}$\tnote{b} & $\gamma_{e^{+}}^{\mathrm{inj}}$ & $C_{\mathrm{s}}^{\mathrm{inj}}$\tnote{b} & $\gamma_{\mathrm{s}}^{\mathrm{inj}}$ & $E^{\mathrm{inj}}_{\mathrm{cut}}$ & $D_{0}$\tnote{c} & $b_{0}$\tnote{d} & $h$\tnote{e} & $H$\tnote{e} & $\phi$\tnote{f} \\
\cline{2-18}
\texttt{P1} & $6.04 \times 10^{41}$ & 3.05 & & & 41.4 & 2.63 & $1.03 \times 10^{41}$ & 3.72 & $1.15 \times 10^{39}$ & 2.08 & $2.80\times 10^{3}$ & 153 & 8.67 & 0.242 & 3.25 & 1.28\\
\hline
Model & $C^{\mathrm{inj}}_{e^{-}}$\tnote{b} & $\gamma_{e^{-}}^{1,\mathrm{inj}}$ & & & $E_{\mathrm{br}}^{\mathrm{inj}}$ & $\gamma_{e^{-}}^{2,\mathrm{inj}}$ & $C^{\mathrm{inj}}_{e^{+}}$\tnote{b} & $\gamma_{e^{+}}^{\mathrm{inj}}$ & $C_{\mathrm{s}}^{\mathrm{inj}}$\tnote{b} & $\gamma_{\mathrm{s}}^{\mathrm{inj}}$ & $E^{\mathrm{inj}}_{\mathrm{cut}}$ & $D_{0}$\tnote{c} & $b_{0}$\tnote{d} & $h$\tnote{e} & $H$\tnote{e} & $\phi_{e^{-}}$\tnote{f} & $\phi_{e^{+}}$\tnote{f}\\
\cline{2-18}
\texttt{P2} & $7.25\times 10^{41}$ & 3.08 & & & 39.3 & 2.66 & $3.61\times 10^{40}$ & 3.22 & $3.54\times 10^{38}$ & 1.84 & $1.68\times 10^{3}$ & 166 & 7.04 & 0.205 & 3.60 & 1.30 & 1.02\\
\hline
\end{tabular}
\begin{tablenotes}
\item[a]{with units of $\mathrm{s^{-1}\,sr^{-1}\,m^{-2}\,GeV^{-1}}$}
\item[b]{with units of $\mathrm{kyr^{-1}\,pc^{-3}\,GeV^{-1}}$}
\item[c]{with units of $\mathrm{pc}^{2}\,\mathrm{kyr}^{-1}$}
\item[d]{with units of $10^{-6}\,\mathrm{GeV}^{-1}\,\mathrm{kyr}^{-1}$}
\item[e]{with units of kpc}
\item[f]{with units of GV}
\end{tablenotes}
\end{threeparttable}
\end{table}
\end{landscape}

\begin{table}
\centering
\caption{A power law injection model.\label{tab:fitprop}}
\begin{threeparttable}
\begin{tabular}{cccccc}
\hline
$D_{0}$\tnote{$\dagger$} & $\delta$ & $b_{0}$ & $\gamma$ & $K$\\
$\left[\mathrm{pc^{2}\,kyr^{-1}}\right]$ & & $\left[\mathrm{GeV^{-1}\,kyr^{-1}}\right]$ & & $\left[\mathrm{kyr^{-1}\,pc^{-3}\,GeV^{-1}}\right]$\\
\hline
100 & $\frac{1}{3}$ & $5 \times 10^{-6}$ & 2.0 & $1.0 \times 10^{39}$\\
\hline
\end{tabular}
\begin{tablenotes}
\item[$\dagger$]{100 $\mathrm{{pc}^{2}\,kyr^{-1}}$$\simeq$ 3$\times$$10^{28}$ $\mathrm{{cm}^{2}\,s^{-1}}$}
\end{tablenotes}
\end{threeparttable}
\end{table}

\section{The injection spectra of cosmic ray electrons and positrons} \label{sec:inj}

The injection spectra of CR $e^{-}$ and $e^{+}$ can be obtained with the above propagation model. To fit the observed spectra by adjusting the injection parameters, we adopt $\delta = 1/3$. We note $\gamma_{e^{-}}^{2} - \gamma_{e^{-}}^{1}$=0.29$\approx (1-\delta)/2$ = 1/3, which motivates us to consider a double power law injection spectrum for the primary CR electrons ($J^{-,\mathrm{inj}}$). The injection spectrum of secondary positrons ($J^{+,\mathrm{inj}}$) and the common component of electrons and positrons ($J^{\mathrm{s,inj}}$) are modeled as a power law and a power law with an exponential cutoff, respectively, in accord with Equation (\ref{eq:eqfit}). Hence, the injection spectra are 
\begin{equation}
\begin{aligned}
J_{e^{-}}^{\mathrm{inj}}&=J^{\mathrm{s,inj}}+J^{-,\mathrm{inj}}+0.6J^{+,\mathrm{inj}}\\
J_{e^{+}}^{\mathrm{inj}}&=J^{\mathrm{s,inj}}+J^{+,\mathrm{inj}},
\end{aligned}
\end{equation}
where, 
\begin{equation}
\begin{aligned}
J^{-,\mathrm{inj}}&=C_{e^{-}}^{\mathrm{inj}}
\begin{cases}
E^{-\gamma_{e^{-}}^{1,\mathrm{inj}}}, \quad &E\leq E_{\mathrm{br}}^{\mathrm{inj}} \\
\left(E_{\mathrm{br}}^{\mathrm{inj}}\right)^{\gamma_{e^{-}}^{2,\mathrm{inj}}-\gamma_{e^{-}}^{1,\mathrm{inj}}}E^{-\gamma_{e^{-}}^{2,\mathrm{inj}}}, \quad &E>E_{\mathrm{br}}^{\mathrm{inj}}
\end{cases} \\
J^{+,\mathrm{inj}}&=C_{e^{+}}^{\mathrm{inj}}E^{-\gamma_{e^{+}}^{\mathrm{inj}}}\\
J^{\mathrm{s,inj}}&=C_{\mathrm{s}}^{\mathrm{inj}}E^{-\gamma_{\mathrm{s}}^{\mathrm{inj}}}\exp\left(-E/E_{\mathrm{cut}}^{\mathrm{inj}}\right).
\end{aligned}
\end{equation}
The primary electron injection spectrum is normalized by $C_{e^{-}}^{\mathrm{inj}}$, whose break energy is $E_{\mathrm{br}}^{\mathrm{inj}}$ and spectral indices are $\gamma_{e^{-}}^{1,\mathrm{inj}}$ and $\gamma_{e^{-}}^{2,\mathrm{inj}}$. The secondary positron spectrum with spectral index $\gamma_{e^{+}}^{\mathrm{inj}}$ is normalized by $C_{e^{+}}^{\mathrm{inj}}$. The injection spectrum of the common component of electrons and positrons is normalized by $C_{\mathrm{s}}^{\mathrm{inj}}$, whose spectral index is $\gamma_{\mathrm{s}}^{\mathrm{inj}}$ and cutoff energy is $E_{\mathrm{cut}}^{\mathrm{inj}}$. \{$D_{0}$, $b_{0}$, $h$, $H$, $\phi$\} are also free parameters. The best-fit parameters are listed in Table \ref{tab:pars}. The equivalent energy density of background photons for the energy loss is about $2.70\,\mathrm{eV\,cm^{-3}}$.

\begin{figure*}
   \centering
   \subfloat[]{\includegraphics[width=0.33\textwidth]{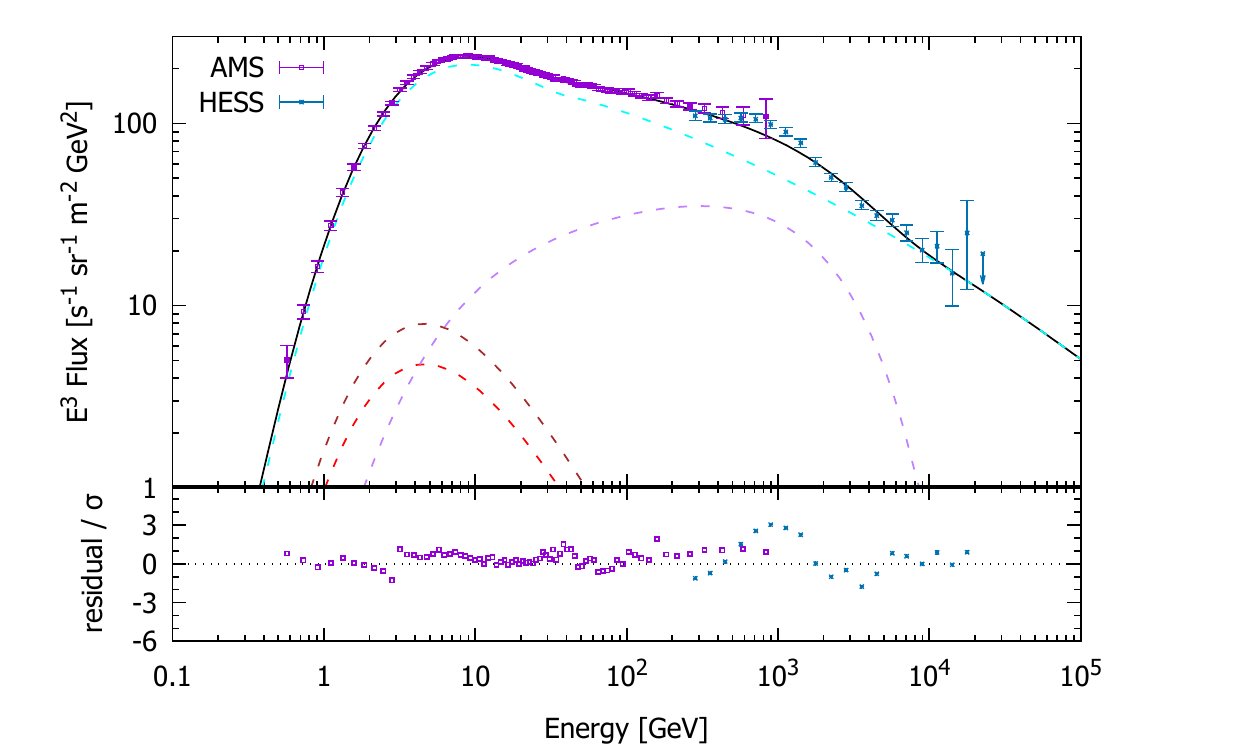}}
   \subfloat[]{\includegraphics[width=0.33\textwidth]{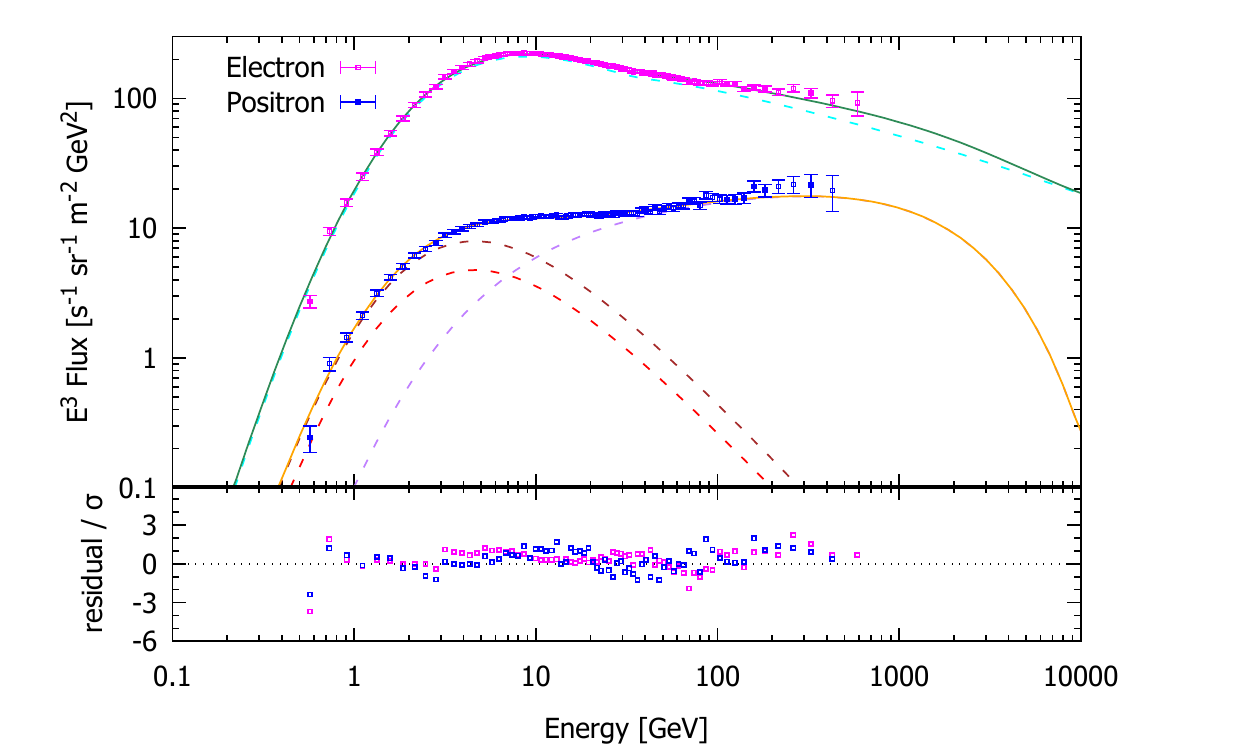}}
   \subfloat[]{\includegraphics[width=0.33\textwidth]{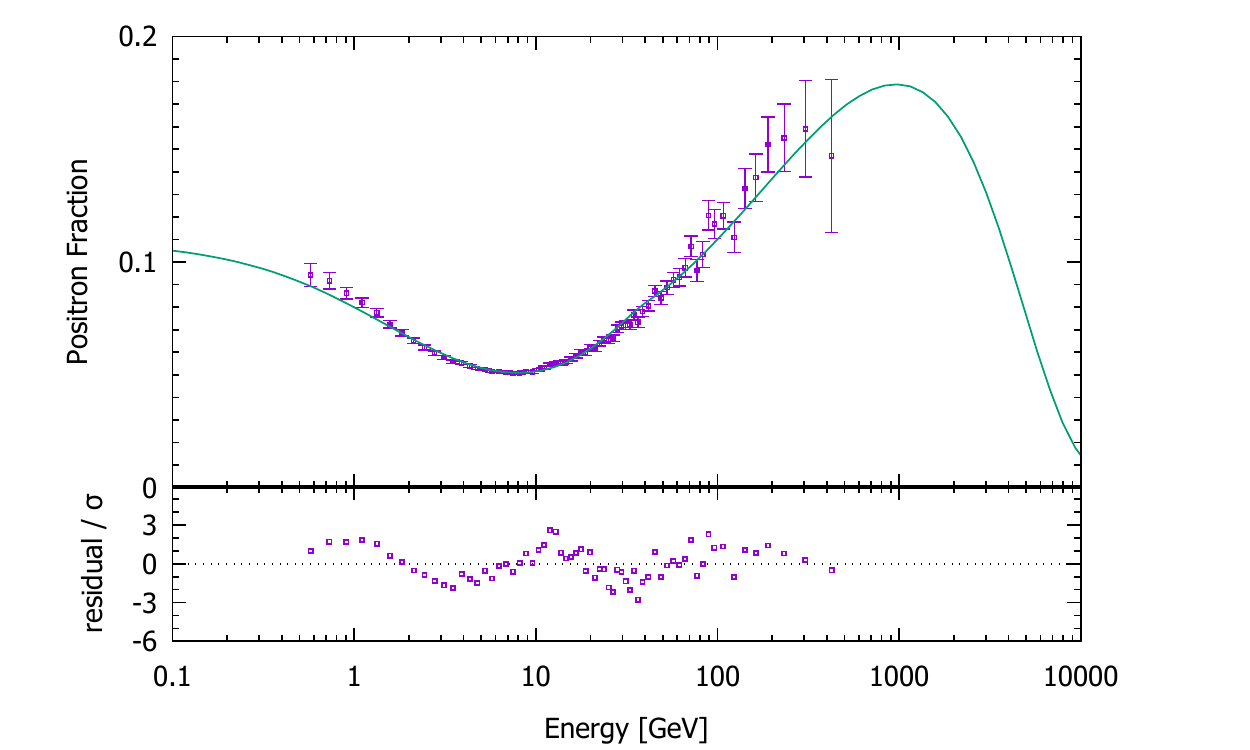}}
   \caption{The fit results according to our propagation model \texttt{P1} with one effective potential. (a): CR $e^{-} + e^{+}$. The cyan dashed line is the flux of primary $e^{-}$, the brown and red dashed lines are the fluxes of secondary $e^{+}$ and $e^{-}$ respectively, the purple dashed line is the common $e^{-}+e^{+}$ flux, and the black line is the sum of the above four components. (b): CR $e^{-}$ and $e^{+}$. The dashed lines are the same as in (a) except that both the fluxes of common $e^{-}$ and $e^{+}$ are shown as purple dashed line. The sea-green and orange lines are the total fluxes of $e^{-}$ and $e^{+}$, respectively. (c): The positron fraction. The ratio of the residual to the standard deviation $\sigma$ of the data are shown in the lower panels. \label{fig:inj}}
\end{figure*}

Figure \ref{fig:inj} shows our best fit (we denote this model by \texttt{P1}) to the spectra of $e^{-}+e^{+}$ (a), $e^{-}$ (b), $e^{+}$ (b), and positron fraction (c). From the best-fit parameters, we have $E_{H}=3.97\,\mathrm{GeV} \approx E_{\mathrm{br}1}=4.96\,\mathrm{GeV}$ implying that the first break of the observed CR $e^{-}$ spectrum results from propagation effects. $E_{h}=9.61\,\mathrm{TeV}$ above which the CR $e^{-}$ spectrum will be softer according to our propagation model and its spectral index is $\gamma_{e^{-}}^{2,\mathrm{inj}}+1=3.63$. The spectral softening above 10 TeV may be tested with future observations such as LHAASO \citep{2016NPPP..279..166D}.

\begin{figure*}
   \centering
   \subfloat[]{\includegraphics[width=0.33\textwidth]{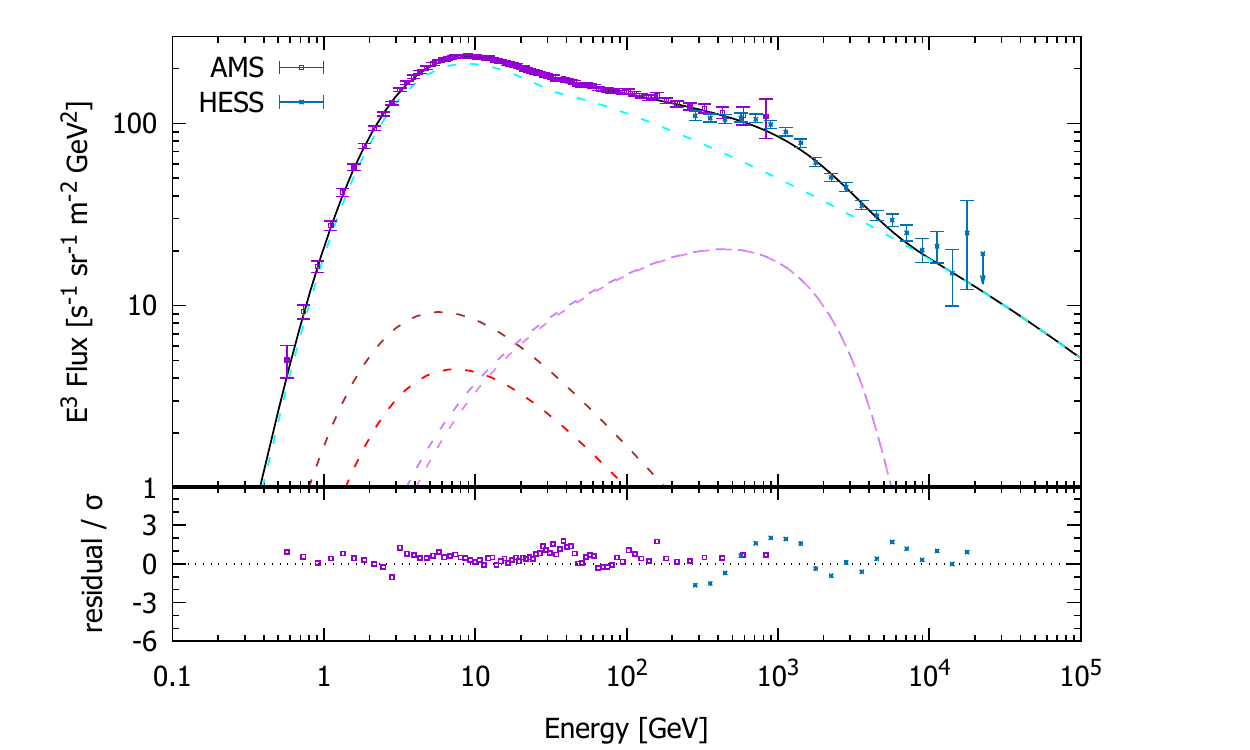}}
   \subfloat[]{\includegraphics[width=0.33\textwidth]{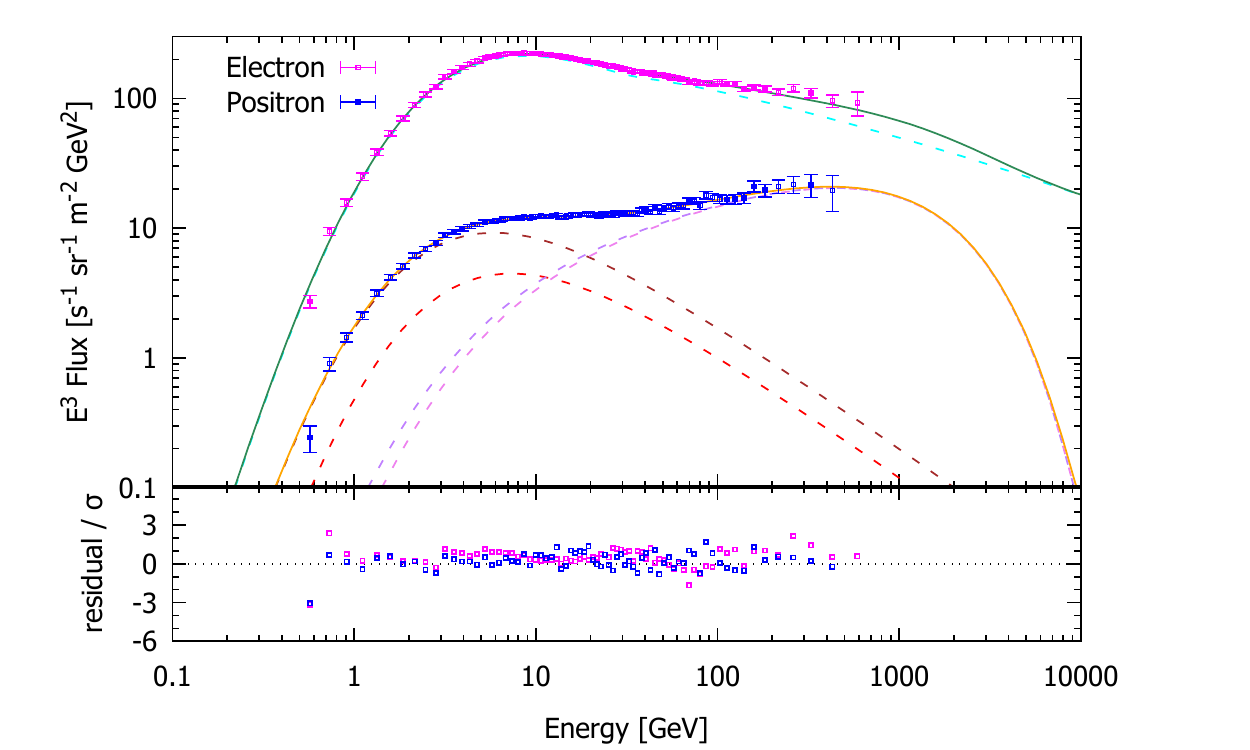}}
   \subfloat[]{\includegraphics[width=0.33\textwidth]{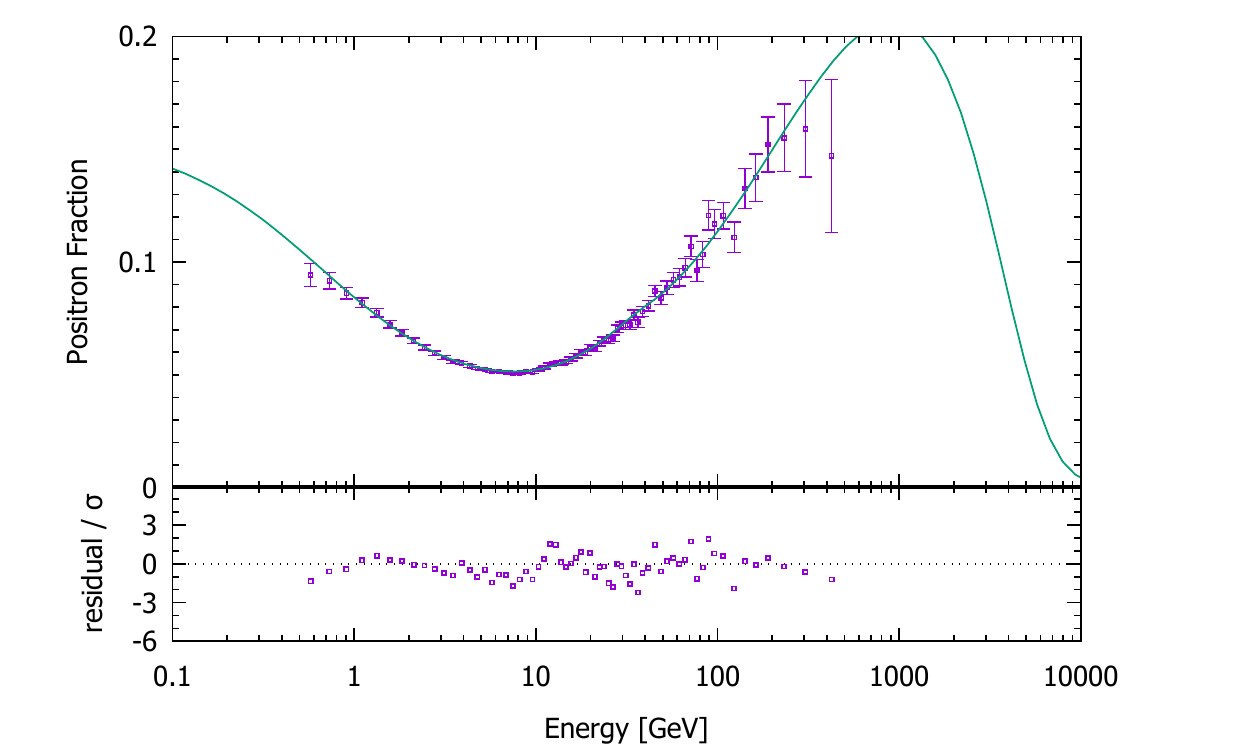}}
   \caption{The fit results according to our propagation model \texttt{P2} with two effective potentials. (a): CR $e^{-} + e^{+}$. The cyan dashed line is the flux of primary $e^{-}$; the brown and red dashed lines are the fluxes of secondary $e^{+}$ and $e^{-}$, respectively; the pink and purple dashed line are the fluxes of common $e^{-}$ and $e^{+}$ flux, respectively; and the black line is the sum of the above five components. (b): CR $e^{-}$ and $e^{+}$. The dashed lines are the same as in (a). The sea-green and orange lines are the total fluxes of $e^{-}$ and $e^{+}$, respectively. (c): The positron fraction. The ratio of the residual to the standard deviation $\sigma$ of the data are shown in the lower panels. \label{fig:inj2p}}
\end{figure*}

Our best fit to the positron fraction according to the above propagation model \texttt{P1} is not as well as the phenomenological model \texttt{M} of Section \ref{sec:spect}, which may be attributed to the difference in the effective potentials of electrons and positrons \citep{2013PhRvL.110h1101M}. Therefore, we also consider two different effective potentials in our propagation model: $\phi_{e^{-}}$ for electrons and $\phi_{e^{+}}$ for positrons. Figure \ref{fig:inj2p} shows such a best fit (we denote this model by \texttt{P2}) to the spectra of $e^{-}+e^{+}$ (a), $e^{-}$ (b), $e^{+}$ (b), and positron fraction (c). The best-fit parameters of model \texttt{P2} are listed in Table \ref{tab:pars}. In Table \ref{tab:chisq}, we list the $\chi^{2}$-values of $e^{-}$, $e^{+}$, $e^{-}+e^{+}$, and positron fraction for model \texttt{P1} and \texttt{P2} (also model \texttt{M}), and we find that model \texttt{P2} gives a much better fit than model \texttt{P1} and has a lower reduced $\chi^{2}_{\mathrm{tot}}$. For model \texttt{P2}, we have $E_{H}$ = 4.51 GeV and $E_{h}$ = 24.4 TeV so that our conclusions drawn for model \texttt{P1} also hold for model \texttt{P2}.

The common component $J^{{\mathrm{s,inj}}}$ has a spectral index $\gamma_{\mathrm{s}}^{\mathrm{inj}}=2.08$ and an exponential cutoff at $E_{\mathrm{cut}}^{\mathrm{inj}}=2.80\,\mathrm{TeV}$ for model \texttt{P1}, while it has $\gamma_{\mathrm{s}}^{\mathrm{inj}}=1.84$ and $E_{\mathrm{cut}}^{\mathrm{inj}}=1.68\,\mathrm{TeV}$ for model \texttt{P2} (see Table \ref{tab:pars}). The spectral indices are very close to the typical ones of pulsar wind nebulae (PWNe) \citep{10.3847/1538-4357/aa7556, 2018A&A...612A...2H}, suggesting that the common component can be interpreted as a continuous distribution of pulsars on the galactic disk.

\begin{table*}
\caption{The chi-square $\chi^{2}$ of model \texttt{M}, \texttt{P1} and \texttt{P2}. From column 2 to 5, we list the $\chi^{2}$-values of $e^{-}$, $e^{+}$, $e^{-}+e^{+}$, and positron fraction, while $\chi^{2}_{\mathrm{tot}}$ is their summation. \label{tab:chisq}}
\begin{tabular}{*{8}{c}}
\hline
Model & $\chi^{2}_{e^{-}}$ & $\chi^{2}_{e^{+}}$ & $\chi^{2}_{e^{-}+e^{+}}$ & $\chi^{2}_{\text{positron fraction}}$ & $\chi^{2}_{\mathrm{tot}}$ & free parameters & reduced $\chi^{2}_{\mathrm{tot}}$ \\
\hline
\texttt{M} & 48.3 & 33.2 & 41.2 & 62.0 & 184.7 & 12 & 0.63 \\
\hline
\texttt{P1} & 54.6 & 52.4 & 73.0 & 97.3 & 277.3 & 14 & 0.96 \\
\hline
\texttt{P2} & 56.1 & 36.0 & 62.0 & 56.0 & 210.1 & 15 & 0.73 \\
\hline 
\end{tabular}
\end{table*}

\section{Conclusion and Discussion} \label{sec:conc}

In this paper, we proposed a parametrized model for the observed fluxes of CR $e^{-}$ and $e^{+}$, and an electron/positron propagation model in the Galaxy. By fitting the spectra of CR $e^{-}$, $e^{+}$, and $e^{-}+e^{+}$, along with the positron fraction, we find that the electron/positron excess above 10 GeV may be attributed to a common power-law component with a high-energy cutoff near 1 TeV and the primary CR electron spectrum has two breaks near 5 GeV and 30 GeV, respectively. For reasonable propagation parameters, we find the spectral break near 5 GeV may be attributed to propagation effects and we expect a spectral softening above $\sim$10 TeV which may be validated with future observations. The injection spectrum of primary electrons is soft at low energies and hardens above $\sim$40 GeV, reminiscence of the ion spectral hardening above 200 GV. If high-energy CRs are mostly accelerated in young supernova remnants (SNRs) as proposed by \citet{2017ApJ...844L...3Z}, our results show that radiative energy loss may affect the spectra of electrons injected by SNRs into the Galaxy significantly. Moreover electron acceleration may be more efficient in young SNRs so that its high energy component is more prominent than those of ions. At very low energies, we adopt a relatively high value of effective potential for the solar modulation. With a relatively low value, \citet{2012ApJ...761..133Y} found that the injection spectrum should become harder below about 5 GeV \citep[see also][]{2012PhRvD..85d3507L,2011A&A...534A..54S}. Better understanding of the effects of solar modulation on electron and position spectra is needed to clarify this issue.

Although the electron/positron excess above 10 GeV and the TeV break of the $e^{-}+e^{+}$ spectrum may be attributed to an identical electron/positron component, their nature remains obscure. PWNe have been considered as dominant contributors to electron-positron excess since the work of \citet{1970ApJ...162L.181S}. Many efforts have been made to understand contributions of nearby PWNe \citep[see, e.g.][and references therein]{2017ApJ...845..107D, 2017ApJ...836..172F} to CR $e^{-}$ and $e^{+}$ fluxes. However, for explaining electron-positron excess our model requires uniformly distributed sources which steadily inject into galactic disk electron-positron. Therefore, besides PWNe due to the time-dependent properties of nearby ones, millisecond pulsars (MSPs) and low mass X-ray binaries (LMXBs) may also be important sources for the stationary common electron/positron component. There are systematic residuals near the 1 TeV break energy of the $e^{-} + e^{+}$ spectrum, which may be improved by adjusting the injection spectrum or by considering contributions from nearby sources.

MSPs are the oldest population of pulsars and have low surface magnetic fields ($\sim$$10^{8}\,\mathrm{G}$). MSPs used to be considered as pair-starved, but the discoveries of a large number of $\gamma$-ray MSPs by Fermi-LAT \citep{2013ApJS..208...17A} changed this picture \citep[see][for more discussions]{2015ApJ...807..130V}. Recently, \citet{2015ApJ...807..130V} accessed contributions of MSPs to the CR $e^{-}$ and $e^{+}$ fluxes by directly calculating realistic source spectra and found a fraction of positron excess can originate from MSPs. The old age and large numbers of MSPs make them promising candidates for our common electron/positron component.

511 keV line emission results from electron-positron annihilation and can be used to map the galactic sources of positrons. INTEGRAL observations \citep{2008Natur.451..159W} indicates that low mass X-ray binaries (LMXBs) may be the dominant contributor to low-energy positrons in the Galaxy. A fraction of LMXBs are microquasars launching jets. For example, V404 Cygni is a microquasar and also a LMXB, where positron annihilation signatures associated with its outburst are found \citep{2016Natur.531..341S}. \citet{2014MNRAS.441.3122G} considered contributions of microquasar jets to the positron excess. Their rough model can explain the rise in the spectrum of CR $e^{+}$ above 30 GeV.

\begin{figure}
   \includegraphics[width=\columnwidth]{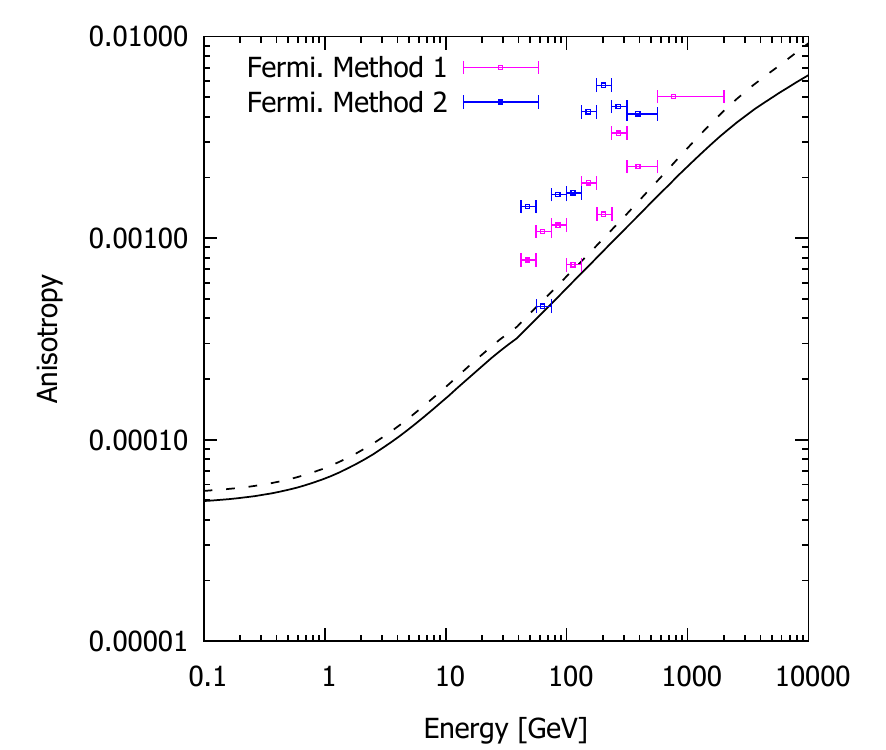}
   \caption{Dipole anisotropy of CR $e^{-}+e^{+}$. The solid and dashed black lines are the predictions according to our model \texttt{P1} and \texttt{P2}, respectively. The upper limits of Fermi-LAT dipole anisotropy measurement are obtained with two different methods \citep{2017PhRvL.118i1103A}. \label{fig:anisot}}
\end{figure}

For nearby source models, the CR flux is expected to enhance in the direction of the sources. Our steady-state model predicts a flux enhancement toward the galactic disk along the local magnetic field line. The dipole anisotropy of CR leptons can be used to distinguish these models \citep{2013ApJ...772...18L, 2017JCAP...01..006M}. Recently, Fermi-LAT collaboration \citep{2017PhRvL.118i1103A} presented upper limits on the dipole anisotropy of CR $e^{-} + e^{+}$ using seven years of data with energies above 42 GeV. Considering the fact that the Earth is not exactly located in the galactic plane with $z_{\earth}$ = 17 pc \citep{10.1093/mnras/stw2772}, we calculate the dipole anisotropy of the model. In the context of diffusive propagation, the dipole anisotropy predicted by our model is given by \citep{2016PhRvL.117o1103A}
\begin{equation}
\Delta_{e^{-}+e^{+}} = \frac{3D}{c}\frac{1}{N}\left|{\partial N \over \partial z}\right|,
\end{equation}
where $N$ is the number density per unit energy of CR $e^{-}+e^{+}$. The solid and dashed black lines of Figure \ref{fig:anisot} show the results for the best-fit model \texttt{P1} and \texttt{P2} described in Section \ref{sec:inj}, respectively. Improved anisotropy measurement can be used to test these models as well.

\section*{Acknowledgements}

This work is partially supported by National Key R\&D Program of China: 2018YFA0404203, NSFC grants: U1738122, 11761131007 and by the International Partnership Program of Chinese Academy of Sciences, Grant No. 114332KYSB20170008.





\begin{thebibliography}{99}

\bibitem[Abdalla et al.(2018)]{2018A&A...612A...2H} Abdalla H. et al.,\ 2018, \href{http://adsabs.harvard.edu/abs/2018A\%26A...612A...2H}{{\color{magenta} \aap}}, 612, A2
\bibitem[Abdo et al.(2013)]{2013ApJS..208...17A} 
Abdo A.~A. et al.,\ 2013, \href{http://adsabs.harvard.edu/abs/2013ApJS..208...17A}{{\color{magenta} \apjs}}, 208, 17
\bibitem[Abdollahi et al.(2017)]{2017PhRvL.118i1103A} Abdollahi S. et al.,\ 2017, \href{http://adsabs.harvard.edu/abs/2017PhRvL.118i1103A}{{\color{magenta} \prl}}, 118, 091103
\bibitem[Abdollahi et al.(2017)]{2017PhRvD..95h2007A} 
Abdollahi S. et al.,\ 2017, \href{http://adsabs.harvard.edu/cgi-bin/nph-basic_connect}{{\color{magenta} \prd}}, 95, 082007
\bibitem[Abeysekara et al.(2017)]{10.3847/1538-4357/aa7556} Abeysekara A. U. et al.,\ 2017, \href{http://adsabs.harvard.edu/abs/2017ApJ...843...40A}{{\color{magenta} \apj}}, 843, 40
\bibitem[Accardo et al.(2014)]{2014PhRvL.113l1101A} Accardo L. et al.,\ 2014, \href{http://adsabs.harvard.edu/abs/2014PhRvL.113l1101A}{{\color{magenta} \prl}}, 113, 121101
\bibitem[Adriani et al.(2009)]{2009Natur.458..607A} Adriani O. et al.,\ 2009, \href{http://adsabs.harvard.edu/abs/2009Natur.458..607A}{{\color{magenta} \nat}}, 458, 607 
\bibitem[Adriani et al.(2011)]{2011PhRvL.106t1101A} Adriani O. et al.,\ 2011, \href{http://adsabs.harvard.edu/abs/2011PhRvL.106t1101A}{{\color{magenta} \prl}}, 106, 201101
\bibitem[Adriani et al.(2013)]{2013PhRvL.111h1102A} 
Adriani O. et al.,\ 2013, \href{http://adsabs.harvard.edu/abs/2013PhRvL.111h1102A}{{\color{magenta} \prl}}, 111, 081102
\bibitem[Adriani et al.(2017)]{2017PhRvL.119r1101A} 
Adriani O. et al.,\ 2017, \href{http://adsabs.harvard.edu/abs/2017PhRvL.119r1101A}{{\color{magenta} \prl}}, 119, 181101
\bibitem[Aguilar et al.(2014a)]{2014PhRvL.113l1102A} 
Aguilar M. et al.,\ 2014a, \href{http://adsabs.harvard.edu/abs/2014PhRvL.113l1102A}{{\color{magenta} \prl}}, 113, 121102
\bibitem[Aguilar et al.(2014b)]{2014PhRvL.113v1102A} 
Aguilar M. et al.,\ 2014b, \href{http://adsabs.harvard.edu/abs/2014PhRvL.113v1102A}{{\color{magenta} \prl}}, 113, 221102 
\bibitem[Aharonian et al.(1995)]{1995A&A...294L..41A} 
Aharonian F.~A., Atoyan A.~M., V\"{o}lk H.~J.,\ 1995, \href{http://adsabs.harvard.edu/abs/1995A\%26A...294L..41A}{{\color{magenta} \aap}}, 294, L41
\bibitem[Ahlers(2016)]{2016PhRvL.117o1103A} Ahlers M.,\ 2016, \href{http://adsabs.harvard.edu/abs/2016PhRvL.117o1103A}{{\color{magenta} \prl}}, 117, 151103
\bibitem[Ambrosi et al.(2017)]{2017Natur.552...63D} Ambrosi G. et al.,\ 2017, \href{http://adsabs.harvard.edu/abs/2017Natur.552...63D}{{\color{magenta} \nat}}, 552, 63
\bibitem[Atoyan et al.(1995)]{1995PhRvD..52.3265A} Atoyan A.~M., Aharonian F.~A., V\"{o}lk H.~J.,\ 1995, \href{http://adsabs.harvard.edu/abs/1995PhRvD..52.3265A}{\color{magenta} \prd}, 52, 3265
\bibitem[Bergstr\"{o}m et al.(2009)]{2009PhRvL.103c1103B} Bergstr\"{o}m, L., Edsj\"{o}, J., Zaharijas, G.,\ 2009, \href{http://adsabs.harvard.edu/abs/2009PhRvL.103c1103B}{\color{magenta} \prl}, 103, 031103
\bibitem[Bulanov \& Dogel(1974)]{1974Ap&SS..29..305B} Bulanov S.~V., Dogel V.~A.,\ 1974, \href{http://adsabs.harvard.edu/abs/1974Ap\%26SS..29..305B}{{\color{magenta} \apss}}, 29, 305
\bibitem[Chang et al.(2008)]{2008Natur.456..362C} Chang J. et al.,\ 2008, \href{http://adsabs.harvard.edu/abs/2008Natur.456..362C}{{\color{magenta} \nat}}, 456, 362
\bibitem[Cholis \& Hooper(2013)]{2013PhRvD..88b3013C} Cholis I., Hooper D.,\ 2013, \href{http://adsabs.harvard.edu/abs/2013PhRvD..88b3013C}{\color{magenta} \prd}, 88, 023013
\bibitem[Delahaye et al.(2009)]{2009A&A...501..821D} 
Delahaye T., Lineros R., Donato F., Fornengo N., Lavalle J., Salati P., Taillet R.,\ 2009, \href{http://adsabs.harvard.edu/abs/2009A\%26A...501..821D}{\color{magenta} \aap}, 501, 821
\bibitem[Delahaye et al.(2010)]{2010A&A...524A..51D} Delahaye T., Lavalle J., Lineros R., Donato F., Fornengo N.,\ 2010, \href{http://adsabs.harvard.edu/abs/2010A\%26A...524A..51D}{\color{magenta} \aap}, 524, A51
\bibitem[Di Mauro et al.(2017)]{2017ApJ...845..107D} Di Mauro M. et al.,\ 2017, \href{http://adsabs.harvard.edu/abs/2017ApJ...845..107D}{{\color{magenta} \apj}}, 845, 107
\bibitem[Di Sciascio et al.(2016)]{2016NPPP..279..166D} Di Sciascio G., LHAASO Collaboration,\ 2016, \href{http://adsabs.harvard.edu/abs/2016NPPP..279..166D}{\color{magenta} NPPP}, 279-281, 166  
\bibitem[Fang et al.(2017)]{2017ApJ...836..172F} 
Fang K., Wang B-B., Bi X-J., Lin S-J., Yin P-F.,\ 2017, \href{http://adsabs.harvard.edu/abs/2017ApJ...836..172F}{{\color{magenta} \apj}}, 836, 172
\bibitem[Farrar(2015)]{10.1017/S1743921316006530} Farrar G.~R.,\ 2015, \href{https://www.cambridge.org/core/journals/proceedings-of-the-international-astronomical-union/article/galactic-magnetic-field-and-its-lensing-of-ultrahigh-energy-and-galactic-cosmic-rays/ABDFB740AC21EFFBEE9524C1D064BBAE#}{{\color{magenta} Proc. IAU}}, 11, 723
\bibitem[Giacalone \& Jokipii(1999)]{1999ApJ...520..204G} 
Giacalone J., Jokipii J.~R.,\ 1999, \href{http://adsabs.harvard.edu/abs/1999ApJ...520..204G}{{\color{magenta} \apj}}, 520, 204
\bibitem[Gleeson \& Axford(1968)]{1968ApJ...154.1011G} 
Gleeson L.~J., Axford W.~I.,\ 1968, \href{http://adsabs.harvard.edu/abs/1968ApJ...154.1011G}{{\color{magenta} \apj}}, 154, 1011
\bibitem[Gupta \& Torres(2014)]{2014MNRAS.441.3122G} Gupta N., Torres D.~F.,\ 2014, \href{http://adsabs.harvard.edu/abs/2014MNRAS.441.3122G}{{\color{magenta} \mnras}}, 441, 3122
\bibitem[Kamae et al.(2006)]{2006ApJ...647..692K} Kamae T., Karlsson N., Mizuno T., Abe T., Koi T.,\ 2006, \href{http://adsabs.harvard.edu/abs/2006ApJ...647..692K}{\color{magenta} \apj}, 647, 692
\bibitem[Karim \& Mamajek(2017)]{10.1093/mnras/stw2772} Karim M.~T., Mamajek E.~E.,\ 2017, \href{http://adsabs.harvard.edu/abs/2017MNRAS.465..472K}{{\color{magenta} \mnras}}, 465, 472
\bibitem[HESS Collaboration(2017)]{HESS2017} Kerszberg D., Kraus M., Kolitzus D., Egberts K., Funk S., Lenain J.-P., Reimer O., Vincent P., 2017, in Proc. 35th Int. Cosmic Ray Conference, Bexco, Busan, Korea., CRI215(ICRC2017)
\bibitem[Kobayashi et al.(2004)]{2004ApJ...601..340K} 
Kobayashi T., Komori Y., Yoshida K., Nishimura J.,\ 2004, \href{http://adsabs.harvard.edu/abs/2004ApJ...601..340K}{{\color{magenta} \apj}}, 601, 340
\bibitem[Linden \& Profumo(2013)]{2013ApJ...772...18L} Linden T., Profumo S.,\ 2013, \href{http://adsabs.harvard.edu/abs/2013ApJ...772...18L}{{\color{magenta} \apj}}, 772, 18
\bibitem[Liu et al.(2012)]{2012PhRvD..85d3507L} Liu J., Yuan Q., Bi X., Li H., Zhang X.,\ 2012, \href{http://adsabs.harvard.edu/abs/2012PhRvD..85d3507L}{{\color{magenta} \prd}}, 85, 043507
\bibitem[Maccione(2013)]{2013PhRvL.110h1101M} Maccione L.,\ 2013, \href{http://adsabs.harvard.edu/abs/2013PhRvL.110h1101M}{{\color{magenta}\prl}}, 110, 081101
\bibitem[Malyshev et al.(2009)]{2009PhRvD..80f3005M} 
Malyshev D., Cholis I., Gelfand J.,\ 2009, \href{http://adsabs.harvard.edu/abs/2009PhRvD..80f3005M}{{\color{magenta}\prd}}, 80, 063005
\bibitem[Manconi et al.(2017)]{2017JCAP...01..006M} Manconi S., Di Mauro M., Donato F.,\ 2017, \href{http://adsabs.harvard.edu/abs/2017JCAP...01..006M}{{\color{magenta} \jcap}}, 01, 006
\bibitem[Moskalenko \& Strong(1998)]{1998ApJ...493..694M} 
Moskalenko I. V., Strong A. W.,\ 1998, \href{http://adsabs.harvard.edu/abs/1998ApJ...493..694M}{\color{magenta} \apj}, 493, 694
\bibitem[Profumo(2012)]{2012CEJPh..10....1P} Profumo S.,\ 2012, \href{http://adsabs.harvard.edu/abs/2012CEJPh..10....1P}{{\color{magenta} CEJPh}}, 10, 1
\bibitem[Schwadron et al.(2014)]{2014Sci...343..988S}
Schwadron N. A. et al.,\ 2014, \href{http://adsabs.harvard.edu/abs/2014Sci...343..988S}{{\color{magenta} \sci}}, 343, 988
\bibitem[Siegert et al.(2016)]{2016Natur.531..341S} 
Siegert T. et al.,\ 2016, \href{http://adsabs.harvard.edu/abs/2016Natur.531..341S}{{\color{magenta} \nat}}, 531, 341
\bibitem[Shen(1970)]{1970ApJ...162L.181S} Shen C.~S.,\ 1970, \href{http://adsabs.harvard.edu/abs/1970ApJ...162L.181S}{{\color{magenta} \apj}}, 162, L181
\bibitem[Strong et al.(2011)]{2011A&A...534A..54S} Strong A. W., Orlando E., Jaffe T. R.,\ 2011, \href{http://adsabs.harvard.edu/abs/2011A\%26A...534A..54S}{{\color{magenta} \aap}}, 534, A54
\bibitem[Venter et al.(2015)]{2015ApJ...807..130V} 
Venter C., Kopp A., Harding A.~K., Gonthier P. L., B\"{u}sching I.,\ 2015, \href{http://adsabs.harvard.edu/abs/2015ApJ...807..130V}{{\color{magenta} \apj}}, 807, 130
\bibitem[Weidenspointner et al.(2008)]{2008Natur.451..159W} 
Weidenspointner G. et al.,\ 2008, \href{http://adsabs.harvard.edu/abs/2008Natur.451..159W}{{\color{magenta} \nat}}, 451, 159
\bibitem[Yuan et al.(2012)]{2012ApJ...761..133Y} Yuan Q., Liu S., Bi X.,\ 2012, \href{http://adsabs.harvard.edu/abs/2012ApJ...761..133Y}{{\color{magenta} \apj}}, 761, 133
\bibitem[Zhang et al.(2017)]{2017ApJ...844L...3Z} Zhang Y., Liu S., Yuan Q.,\ 2017, \href{http://adsabs.harvard.edu/abs/2017ApJ...844L...3Z}{{\color{magenta} \apjl}}, 844, L3

\end{thebibliography}







\bsp	
\label{lastpage}
\end{document}